# XML Technologies in Computer Assisted Learning and Testing Systems

**Adrian Cojocariu, Cristina Ofelia Stanciu**
„Tibiscus" University of Timişoara, Romania

**ABSTRACT**: The learning and assessment activities have undergone major changes due to the development of modern technologies. The computer-assisted learning and testing has proven a number of advantages in the development of modern educational system. The paper suggests a solution for the computer-assisted testing, which uses XML technologies, a solution that could make the basis for developing a learning computer-assisted system.

The new technologies are offering great possibilities regarding education, due to the hardware and software development. One of the main objectives of modern education is the permanent improvement of the teaching and learning processes by using the available multimedia technologies. The multimedia technologies offer a connection between the audio-visual effects in order to develop complex and suggestive courses and presentation. The evolution of the information and communication technologies has also lead to the development of e-learning technologies. Due to the Internet, the web product generating tools, the audio and video recording, there have been elaborated online courses and educational software. By virtual learning we understand e-learning and educational software, and it is considered to be a very efficient and useful way of learning. This kind of education allows the student to chose what, when, where and how much to learn, situations that are according to the new paradigm of education.

It is important to understand the e-learning and educational software concepts and to show their objectives. E-learning contains traditional methods and techniques and by using the information and communication technologies will assist the individual in achieving knowledge and skills in certain fields. It is an accessible way to information and knowledge, and offers new and efficient methods of teaching, learning, permanent education and knowledge evaluation. E-learning technologies may be complementary or an alternative to the traditional education methods. E-learning allows organizing the on-line learning process by subjects or themes, while traditional education is generally organized in groups by age.

By integrating new technologies into methods of education, by conducting researching activities in the areas of standards and the cognitive psychology methods, we develop actions regarding the improvement of the teaching-learning-evaluation activity on any level or form, and in any curricular area.

The teaching-learning-testing process is going through important changes due to the e-learning technologies. The education system is definitely involved in the foundation and building of the information and knowledge society.

The educational software is considered to be any software application that is able to run on a personal computer and that shows a certain topic, theme or practical experience, or a course, being a great alternative to the traditional education methods.

Roger Bohn claims about the learning process that "*Learning is evolution of knowledge over time*", considering the quick evolution of the information and communication technologies. Technology Based Learning is significant when referring to a education forms that is using other education tools that the traditional class-room forms that include computers, television, multimedia machines. Computer Based Learning has proved an important impact upon the knowledge domain.

Lewis Perelman claims that "*We must replace education and schools with hyper-education, that is not only a new form of education, without any constrains but also a world without education constrains. The nations that chose to use the newest education systems based on the newest technologies will be the most powerful nations of the 21th century. This education form will extend beyond school, beyond the static role of the teacher and education and beyond the school years. The intelligent education environments, the interactive hypermedia systems, the biomedical and intelligent technologies, the communication infrastructure are making possible the access to knowledge anytime and anyplace. The HyperLearning revolution impact includes access to education through distance education, intelligent education environment, that are able to adapt to particular* *learning characteristics, any age and level person having access to education."*

Data management is a very important aspect of the learning and testing system, and as the data amount is continuously enlarging, one has to find proper solutions to store and manage data. XML (eXtensible Markup Language) offers an organized and elegant way to store data. Its main advantage is the adaptability, and the fact that data modeled with XML are human readable, which in many cases proves to be a great advantage.

XML is a modern format and most if the visual high level programming languages, such as Visual C++, C#, Java, are capable to process and manage XML files and the data stored in these files.

Modeling data in XML format also offers the possibility to validate data by defining so called XML schemes. The XSD (*XML Schema Definition*) files define rules and patterns the XML file should fit in. The assemble made up by the data models represented in XML format together with the XSD schemes and the applications developed in programming languages that offer function libraries for processing these data models (.Net, Java packages, Qt) represents a powerful and efficient solution, but mostly an elegant one, according to the object oriented programming point of view.

There are two main methods for parsing XML files: SAX and DOM. Each of these methods has its own advantages:

- SAX (*Simple API for XML*) is an event based method: at the parsing moment of each XML entity of the model a signal is output, signal which is received by the software developer that uses the SAX functions and will be able to realize the internal model of the data from the XML file. Among the main particularities of a SAX parser, we can mention the fast processing and simplicity.
- DOM (*Document Object Model*) is based on loading the whole content of the XML file in the memory and on the disposal of its elements to the software developer in the tree model which is typical to the format. A derived advantage of this parsing way is that in any moment one can access a node of the XML file through the functions and data types offered by the DOM function library implemented in a certain programming language. Among the main characteristics of DOM parser we mention the model hierarchy and the easy serialization.

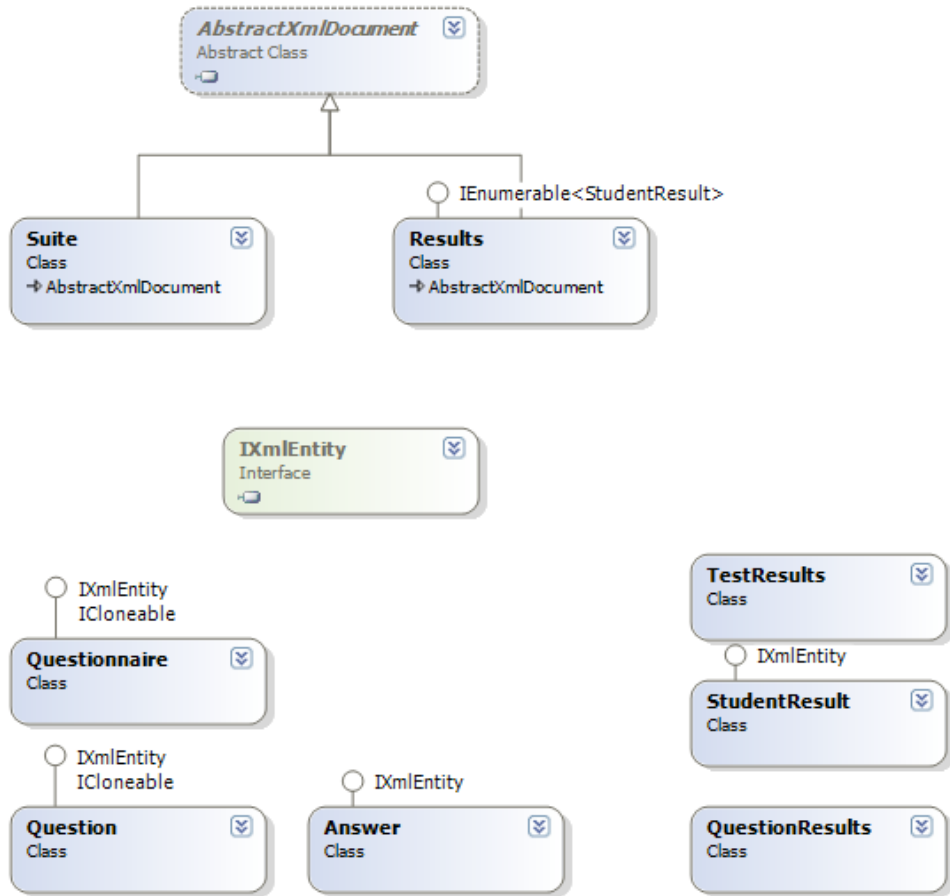

Figure 1. Class diagram for the GrileX application

We have developed a software application for quiz-based testing. The application, called *GrileX* has been developed in the C# programming language, in the Visual Studio .Net 2005 environment. Data is being stored in XML files, and the correct answers are coded with MD5 (Message-Digest algorithm 5) algorithm. This security algorithm is a widely used cryptographic hash function with a 128-bit hash value [wikipedia]. MD5 has been used in a wide variety of security applications, and is also used to check the integrity of different files. The class diagram on which the software application has been developed is presented in Figure 1.

The software application is based on the client-server architecture (Figure 2), and actually a suite that consists in three parts: a Server that runs on the server computer, the Testing application that runs on the client computers and the Administration application which allows configuring the tests. The Administration application can only be accessed by the professors using an ID and a password. The Testing application is the actual testing module, used by the students. Before the beginning of the test they enter their personal data (name, year of study, subject) and at the end of the testing process the result will be shown in a dialogue window. All this data regarding a student is sent to the Server so the professor is aware and supervises the whole testing process.

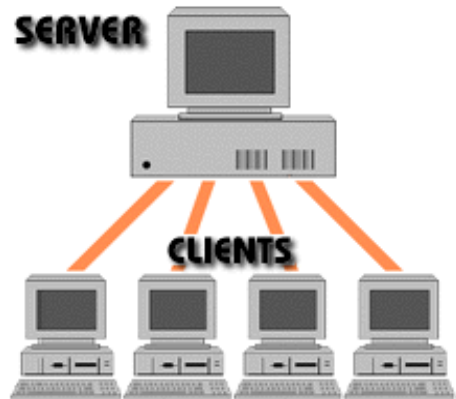

Figure 2. The client-server architecture

So far, computer assisted testing proved to have several advantages, but also disadvantages.

Some of the advantages are:
- the increase of the evaluation process speed;
- higher objectivity;
- the decrease of evaluation errors;
- assures transparency in the examination process;
- diminishes the nervous and emotional state of the student.

Some of the disadvantages are:
- requires appropriate technical support;
- may reduce the capability of verbal communication of the students;
- may lead to the loss of argumentation and discussion capability.

In spite of the above shown disadvantages, the so far developed application has encountered a real success, from students and teachers, and for this reason we are aiming to extend its features and develop a complex learning and testing system that is using new and modern technologies, such as XML.

**References**

[Coj07]  Cojocariu, A., *The knowledge society and the modern university education system*, Anale. Seria Ştiinţe Economice. Timişoara, Vol. XIII, Ed. Eurostampa, 2007

[EY99]  English, S., Yazdani, M., *Computer-supported cooperative learning in a Virtual University*, Journal of Computer Assisted Learning, Vol. 15, No. 1, March 1999

[JMRE02] Joy, M., Muzykantskii, B., Rawles, S., Evans, M., *An Infrastructure for Web-Based Computer-Assisted Learning,* Journal on Educational Resources in Computing (JERIC), 2002

[Roş02]  Rosca, I. Gh., Zamfir, G., *Informatica instruirii*, Editura Economica, 2002

[Tay01]  Taylor, James C., *Fifth generation distance education.* e-Journal of Instructional Science and Technology (e-JIST), 2001

[Vla05]  Vlada, M., *Tehnologiile societăţii informaţionale*, CNIV-2005, Virtual Learning - Virtual Reality, Conferinţa Naţională de Învăţământ Virtual, Software şi Management Educaţional, Ediţia a III-a, Editura Universităţii din Bucureşti, 2005

[Vli02]  van der Vlist, E., *The W3C's Object-Oriented Descriptions for XML*, O'Rellys Publishing, 2002

[Yer06]  Yergeau, F., colab, *Extensible Markup Language 1.0. (Fourth edition),* (http://www.w3.org/TR/REC-xml/ - W3C Recommendation 16 August 2006, edited in place 29 September 2006)